# Facile "Pick-up" experiments and Monte Carlo simulations for the entanglement of tunable staple-like particles

Youhan Sohn, Saeed Pezeshki, and Francois Barthelat*

Department of Mechanical Engineering, University of Colorado, 427 UCB, 1111 Engineering Dr, Boulder, CO 80309, USA

* Correspondence to: francois.barthelat@colorado.edu

## Abstract

Entangled matter provides intriguing perspectives in terms of deformation mechanisms, mechanical properties, assembly and disassembly. However, collective entanglement mechanisms are complex, occur over multiple length scales, and they are not fully understood to this day. In this report, we propose a simple pick-up test to measure the entanglement in staple-like particles with various leg lengths, crown-leg angles, and backbone thickness. We also present a new "throw-bounce-tangle" model based on a 3D geometrical entanglement criterion between two staples, and a Monte Carlo approach to predict the probabilities of entanglement in a bundle of staples. This relatively simple model is computationally efficient and it predicts an average density of entanglement which is consistent with the entanglement strength measured experimentally. Entanglement is very sensitive to the thickness of the backbone of the staples, even in regimes where that thickness is a small fraction (<0.04) of the other dimensions. We demonstrate an interesting use for this model to optimize staple-like particles for maximum entanglement. New designs of tunable "entangled granular metamaterials" can produce attractive combinations of strength, extensibility, and toughness that may soon outperform lightweight engineering materials such as solid foams and lattices.

simulations

# 1 Introduction

Typical granular materials made of spherical or quasi-spherical grains require mechanical confinement to generate shear strength [1-3] or a cohesive second phase at the interface between the grains [4, 5]. Grains with more extreme geometries such as elongated rods can assemble into free standing structures with some tensile strength, because of long range interactions and multiple contact points [6-9]. Long rods, in turn, may be assembled into hexapods [10] or other star-like particles with entanglement or "geometric cohesion" [11], offering intriguing possibilities in terms of structural design and architecture [12, 13]. Even more extreme designs have branches with hooks and barbs, with the classical example of U-shape staple-like particles [14-17]. These particles can indeed latch and hook onto one another, generating substantial tensile strength [15, 18, 19]. Tensile force chains develop in these materials [16], and although they tend to be more sparse than typical compressive chains, they are sufficiently strong and stable to enable free-standing structures (beams, columns [14, 20]). How the shapes of these particles govern entanglement, and in turn translate into strength provides a rich landscape in terms of mechanics and design. For example, geometrical alteration on standard staples, such as changing the length of legs [14] or twisting of the legs [20, 21], have been shown to have profound impact on entanglement and strength, and interestingly, some optimum geometries have already been identified within these design spaces. Various experimental approaches have been used to assess entanglement strength. Perhaps the simplest of these experiments consists of assessing the size of an entangled bundle of particles that can be lifted by just picking up a few staples against gravity [6, 9, 20]. More complex experiments have measured angle of repose [16], and the stability of long free-standing columns [22] and short columns subjected to vibrations [14]. Other mechanical tests on entangled materials have included tensile tests [15, 19] and flexural tests [20]. While these experiments have provided "macroscopic" mechanical properties for bundles of entangled particles, they provide limited insights into the fundamental mechanisms of entanglement and disentanglement at the local level. In addition, each of these experimental approaches was used to explore only one particular aspect of particle design (for example, [14] focused on the effect of leg length only). The lack of "unified" testing methods makes it difficult to assess and compare the entanglement efficacy of staples with different designs. To gain insights into the mechanics of entanglement, numerical models were also developed, primarily based on the discrete element method (DEM) [23]. These models have revealed the effect of packing density on entanglement strength [14], the impact of alignment of staple-like particles across gravity [20], and the dynamic structure of tensile force lines [16, 19, 24]. However, these DEM models, performed on thousands of particles with complex shapes, can be computationally expensive and produce large amounts of data that can be difficult to interpret. In this report, we present a relatively simple experiment to measure entanglement based on the bundle pick-up method. The second part of the report presents an entanglement model based on a pair of staples. A simple geometrical criterion for entanglement, together with a Monte Carlo approach, produces a prediction for entanglement probability and the volumetric density of entangled particles that agree well with experiments on staples with a variety of designs.

## 2 Experiments

The objective was to provide a simple and repeatable protocol to measure the entanglement and strength of various staple-like particle geometries. The base particles we used for this study were standard steel office staples (Swingline, IL), with dimensions shown in Fig. 1a. These staples come in the form of "sticks" of about 200 staples, bonded by a relatively weak polymeric adhesive. To separate individual staples, we immersed sticks of staples in acetone, which immediately dissolved the adhesive and detached the staples. After thorough cleaning and drying, bundles of staples were prepared for the pick-up test. One thousand staples were first loosely poured into a container with the shape of a truncated cone, with a base diameter of 80 mm (approximately 6.3 times the length of the crown of individual staples, Fig. 1b). Next, we used a custom "pick-up" tool made of three staples embedded at the end of a 3D printed handle (Fig. 1c) to engage the staples in the center of the container and to pull them vertically against gravity (Fig. 1d). The number of staples picked up by this process depends on how much entanglement is present in the bundle, which itself is a strong function of the geometry of the individual staples. Some geometries led to poor entanglement and few staples picked up (Fig. 1e), while other geometries generated more entanglement and much larger groups of picked-up staples (Fig. 1f). To quantify the entanglement of staples, we used a "picked-up fraction," which we defined as the fraction of staples lifted from the initial bundle of 1000 staples (measured by weighing the staples picked from the bundle). Repeated tests on the same staples revealed typical variations of 5-10% around the mean value of the picked-up fraction. This repeatability was sufficient to discriminate between different staple geometries.

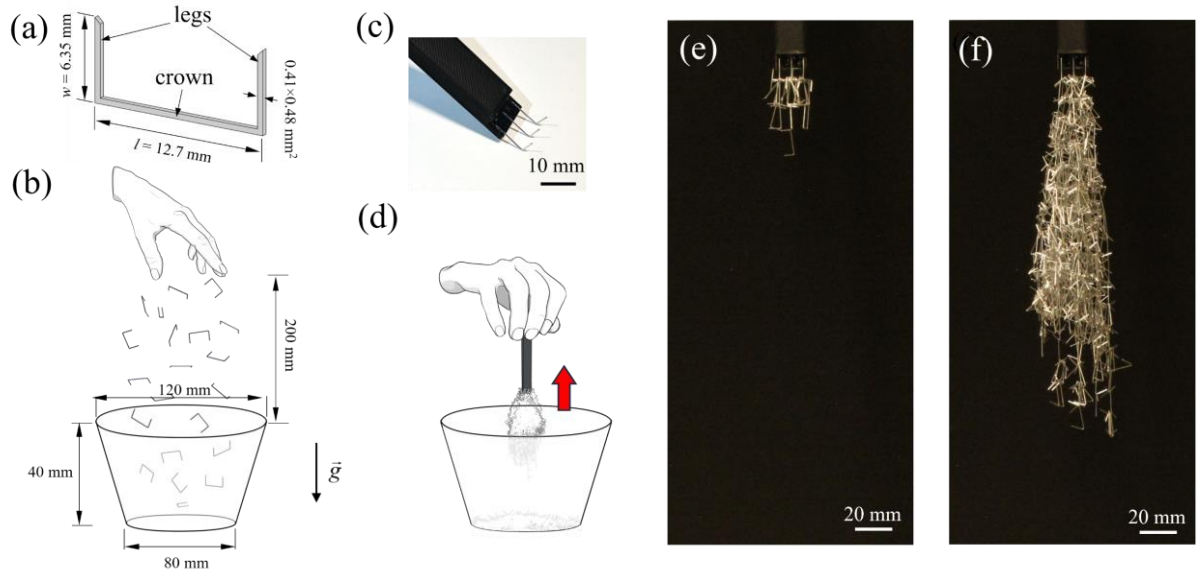

**Fig. 1** Overview of the "pick-up" experiment: **a** Individual standard office staple with dimensions; These staples are **b** poured into an open container. **c** A custom "fishing tool" was used to **d** grab the center staples and pull upwards. The amount of picked-up staples was a strong function of geometry and could be **e** small (a few staples) to **f** large (most of the staples picked up).

We now present the first part of the study, where we varied the angle between the legs and the crown of the individual staples (the crown-leg angle $\theta$). We designed and fabricated 3D-printed tools to fold the legs to either decrease or increase $\theta$ from $\theta = 90°$ reference sticks of staples (Fig. 2a). We then performed three pick-up experiments on each of these geometries. Figure 2b shows the pick-up fraction as a function of the crown-leg angle $\theta$, together with three superimposed contours of the picked-up bundle. The reference staples produced a modest entanglement, with an average pick-up fraction of only about 0.03. As expected, this number was even smaller for high angles ($\theta = 120°$). On the other hand, decreasing $\theta$ greatly improved entanglement, with a pick-up fraction greater than 0.8 for $\theta = 60°$. An intuitive explanation is that decreasing $\theta$ turns the staples into a pair of increasingly sharp "hooks," which can generate more robust entanglement with other staples. However, decreasing the angle further led to poorer entanglement, with a pick-up fraction of less than 2% for $\theta = 20°$. We hypothesized that another effect is at play: The reduction of $\theta$ in effect "closes" the geometry of the staple, reducing the probability of the staples to geometrically "engage" with one another. The observed entanglement peak $\theta = 45\text{-}60°$ would then be the result of two competing mechanisms: closing $\theta$ would make the entanglement between two staples stronger once they engage, but it also decreases the probability of staples mutually engaging. In the following sections, we introduce a simple model that captures the competition of these mechanisms.

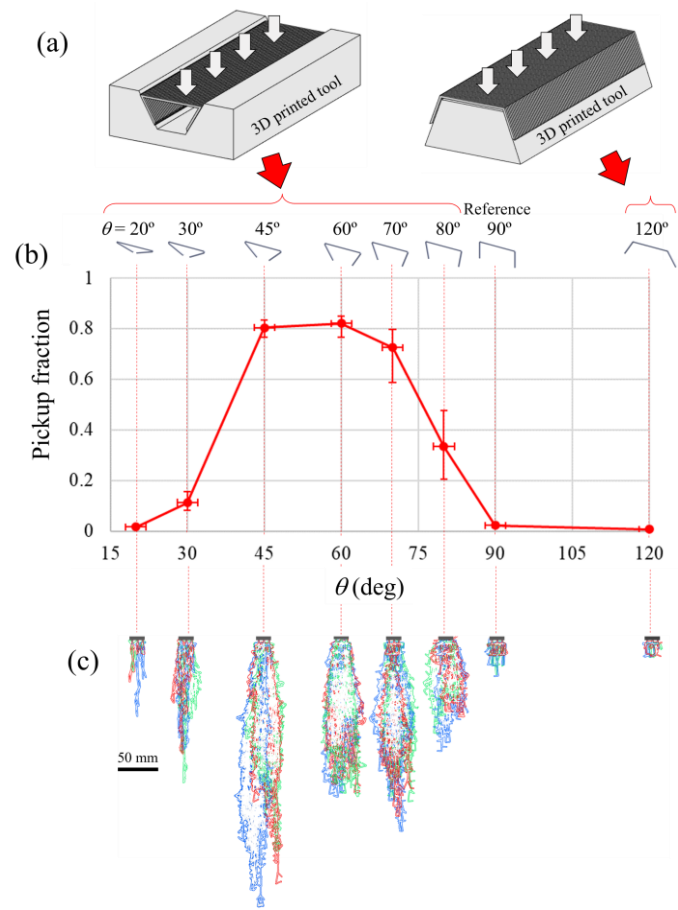

**Fig. 2 a** The crown-leg angle of individual staples can be increased or decreased from the $\theta = 90°$ reference using custom 3D printed embossing tools [19]; **b** Experimental pick-up fraction as a function of crown-leg angle $\theta$; **c** Composite images of the picked-up bundles.

# 3 A "throw-bounce-tangle" model for geometric entanglement

Entanglement and disentanglement are complex processes that involve multiple spatial and time scales [15, 16, 25]. In this study, we sought a relatively simple model to capture entanglement at a fundamental level, i.e., between two staples, based on geometry only. We assumed a round cross section for the backbone of the staples (with diameter $d$ = 0.45 mm for a standard office staple) with semi-spherical caps (Fig. 3a). This simplification streamlined 3D calculations for staple-to-staple distance, the evaluation of collisions, and other staple-staple interactions.

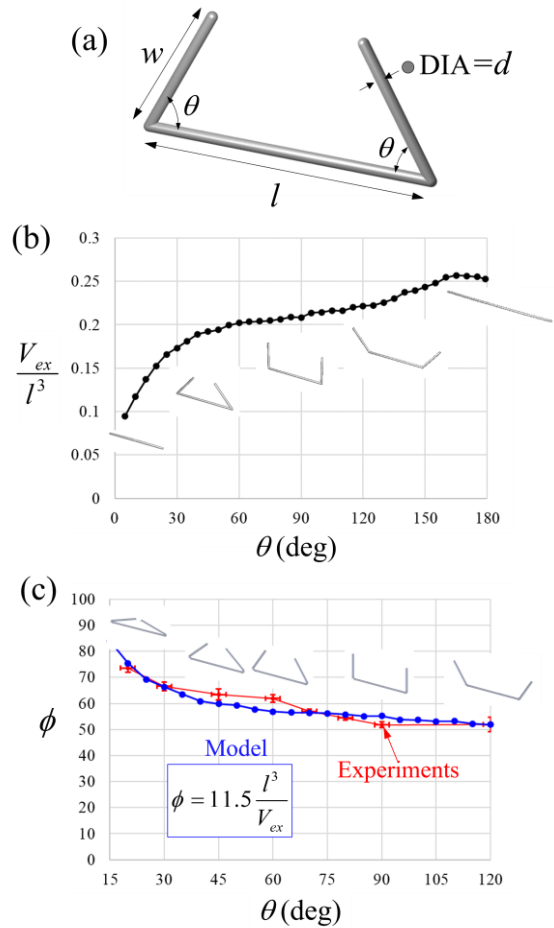

**Fig. 3 a** Geometry of the staples used in the MC models; **b** Nondimensional excluded volume and **c** experimental and modeled volumetric density of staples as a function of crown-leg angle $\theta$.

*Excluded volume and packing fraction*: The excluded volume provides a measure of the density for the staples in a bundle [6, 25]. We computed the excluded volume of individual staples using Monte Carlo simulations, following Gravish et al. [25]. We first considered a center staple (staple 1) and a spherical volume *V* centered on that staple with a radius several times the size of the staple. We then generated a second staple (staple 2) at a random position and random orientation within that volume. The occurrence of collisions between staple 1 and 2 was then computed using 3D geometry accounting for the relative position of staple 1 and 2 and the diameter of the backbone. The probability of collision between the two staples is therefore $c/N$ where *c* is the total number of collisions over *N* realizations. The excluded volume is simply given by:

$$V_{ex} = \frac{c}{N} V \qquad (1)$$

If *N* is sufficiently large ($N > 10^6$ in our simulations), we found that the result $V_{ex}$ is independent of the simulation volume *V*. The model was implemented using MATLAB with parallel processing [26]. Using this model, we recovered the theoretically excluded volume of rods of different aspect ratios [6] and the excluded volume of staples with varying leg lengths [25]. Figure 3b shows the results of the excluded volume of standard office staples ($w/l = 0.5$, $d/l = 0.035$) as a function of crown leg angle $\theta$. Using previous results on random packing of rods [6] and staples [25], the packing factor (*PF*) of a bundle of staples can be written:

$$PF = C \frac{V_p}{V_{ex}} \qquad (2)$$

Where $V_p$ is the volume of the individual staple and *C* is a constant parameter that corresponds to the average number of staples in contact with a given staple. *PF* represents the volume fraction of "solid material" in the bundle. As opposed to porous or cellular material whose properties are a strong function of solid volume fraction, in bundles of staples it is the number of staples per unit volume, or volumetric staple density $\phi$, that is important for strength. It is written:

$$\phi = \frac{l^3}{V_p} PF = C \frac{l^3}{V_{ex}} \quad (3)$$

Note that to keep the results nondimensional, $\phi$ is written as the average number of staples in a $l \times l \times l$ volume (i.e., the unit volume is expressed in unit of $l$). To calibrate the constant $C$, we experimentally measured $\phi$ by pouring 1000 staples into a transparent acrylic container with a round section (diameter = 38 mm). The height of the bundle was used to compute its volume and then to compute $\phi$. Figure 3c shows $\phi$ as function of crown-leg angle $\theta$, showing a decrease from about 75 to 50 as $\theta$ is increased from $\theta = 20°$ to $120°$. Equation (3) was fitted onto that data to produce $C \approx 11.5$. This C value is close, but slightly larger than the value of $C \approx 8.75$ obtained through oscillatory excitation on staples by Gravish et al. [14] and larger than for rods [6].

*Modeling entanglement with the throw-bounce-tangle model*: The aim of the "throw-bounce-tangle" model we present here was to predict the probability of entanglement of a pair of staples-like particles based on their geometry. The model is centered on a staple (staple 1) which remains stationary. Another staple (staple 2) is placed at a random position on a sphere of fixed radius centered on staple 1, and at a random orientation. In this model, the radius of that sphere is to represent the typical distance between staples in the bundle, and for this reason, we used the "excluded radius" computed from the excluded volume:

$$R_{ex} = \sqrt[3]{\frac{3}{4\pi} V_{ex}} \quad (4)$$

Initial conformations where staples 1 and 2 collide are not permitted and were rejected. Next, we considered a "throw" step that simulates pouring or vibrations, which would provide an impulse of displacement to staple 2. We considered all possible 3D directions for a translation of staple 2 from the initial position, which can lead to two outcomes: Either Staple 2 collides with particle 1, or staple 2 "misses" particle 1. In our algorithm, we considered all possible directions for an impulse on staple 2, in the form of a total solid angle $\Omega = 4\pi$. Using simple 3D geometrical rules, we then computed, numerically, which "visibility" solid angle $\Omega^{(v)}$ led staple 2 to geometrically intersect with staple 1. Figure 4a provides a simplified 2D diagram showing the visibility angles $\Omega_A^{(v)}$ and $\Omega_A^{(v)}$ from two possible positions $A$ and $B$ for staple 2. This process of random placement of staple 2 on the excluded sphere and calculation of the visibility solid angle was repeated $N$ times, so that the probability of staple 2 and 1 to interact was given by:

$$p_v = \frac{1}{N}\sum_i^N \frac{\Omega_i^{(v)}}{4\pi} \qquad (5)$$

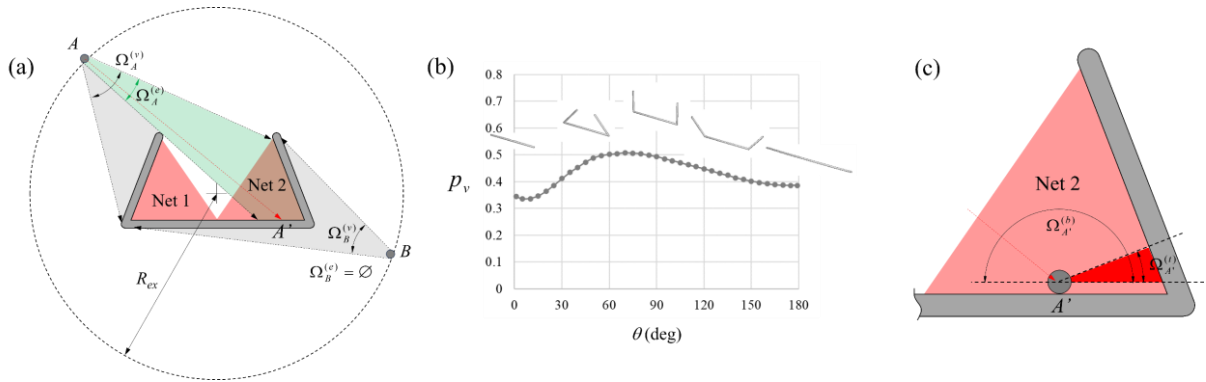

**Fig. 4 a** 2D Diagram showing two examples of possible outcomes for the "throw" stage (the actual model is fully three dimensional); **b** "visibility" $p_v$ of individual staples as a function of crown-leg angle $\theta$; **c** 2D Diagram showing how a bounce is considered, and a possible further entanglement is assessed.

The probability $p_v$ therefore provides a measure of how statistically "visible" each staple is to neighboring staples. A plot of $p_v$ as a function of crown-leg angle $\theta$ (Fig. 4b) shows that $p_v$ is the lowest when legs and crown are aligned, so the staples are effectively rods (cases $\theta = 0°$ and $\theta = 180°$). Next, a director vector was randomly selected from the solid angle sector $\Omega^{(v)}$. We computed the displacement required for staple 2 to contact staple 1 along that director angle, and we updated the positions of staple 1 and 2, now in contact. The interaction of two staples by direct contact does not, however, guarantee that they will entangle. To determine whether the two staples geometrically "engage" as a first step to entanglement, we defined a set of "catch nets" for each staple, as shown in Fig. 4a. The catch nets are plane regions partially enclosed by the crown and the legs of the staples, where entanglement may take place. We then determined whether the two staples "engaged," which we defined as conformations that satisfied the condition of reciprocal engagement: (i) Any branch of staple 2 intersects with a net from staple 1, and (ii) any branch of staple 1 intersects with a net from staple 2. If any of these conditions were not verified, then the particles simply contacted with no engagement. To compute the probability for the two staples to transition from a "free" state to a state of "engagement" $p_e$, we determined numerically which subset $\Omega^e$ of $\Omega^{(v)}$ led to engagement between the staples. Figure 4a illustrates this process in two dimensions, where only a section of the backbone of staple 2 is shown. In this example, if staple 2 has an initial position at $A$, there is a set of directions $\Omega_A^e \in \Omega_A^{(v)}$ where staple 2 can engage with staple 1 through its catch net 2. If staple 2 has an initial position at $B$, there are no possible engagement and $\Omega_B^e = \varnothing$. The probability $p_e$ is then given by:

$$p_e = p_v \frac{1}{N} \sum_i^N \frac{\Omega_i^{(e)}}{\Omega_i^{(v)}} \qquad (6)$$

Figure 5a shows these probabilities for two staples as a function of crown-leg angle (using $N = 10^6$ realizations). The probability of engagement is the highest in the 60-90° range, and it is zero for $\theta = 0°$ and $\theta = 180°$ because the catch nets have zero surfaces (in these two extreme cases, the staples are rods).

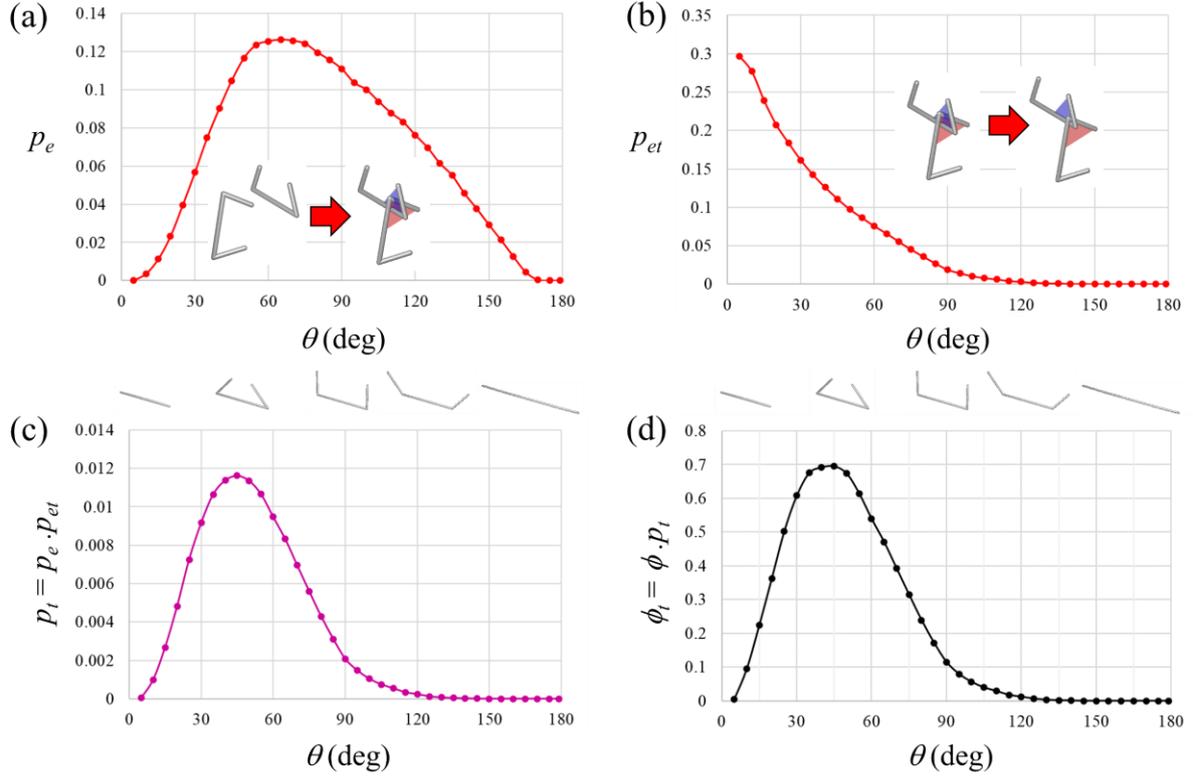

**Fig. 5** Main results from the throw-bounce-tangle model for a pair of staples of different crown-leg angles $\theta$: **a** Probability of engagement; **b** probability of transitioning from "engaged" to "entangled"; **c** probability of entanglement and **d** Volumetric density of entangled staples as a function of crown-leg angle $\theta$.

The final step in the model is a random "bounce" of staple 2 from staple 1 and a calculation of the probability of this bounce to further engage the two staples, in which case the two staples are entangled. This additional step was important to capture the propensity of the particles to keep other particles trapped by entanglement. For this step, we only consider cases where the two staples are already engaged (Fig. 4c). We first defined a solid angular sector $\Omega^b$ defining the possible directions of a bounce away from the contact point. Within these directions, we then geometrically determined which directions $\Omega^{(t)}$ would take staple 2 further toward the corner of the net. The probability $p_{et}$ of transitioning from simple engagement to an entangled conformation is then:

$$p_{et} = \frac{1}{N}\sum_i^N \frac{\Omega_i^{(t)}}{\Omega^{(b)}} \qquad (7)$$

Figure 5b shows the probability of further entanglement from an engaged configuration as a function of crown-leg angle. Once the staples are engaged, the probability of entanglement increases rapidly when the angle $\theta$ is decreased. In other words, staples with smaller $\theta$, if they engage with other staples, are much more likely to geometrically "trap" these staples. The entanglement probability between two staples, accounting for all three possible transition paths to entanglement configurations, can then be written:

$$p_t = p_e \cdot p_{et} \qquad (8)$$

Figure 5b shows the probabilities $p_t$ for two staples are function of crown-leg angle. The model predicts an entanglement probability which is maximum at $\theta \sim 50°$ and which vanishes towards $\theta = 0°$, and $\theta = 180°$. Finally, as pointed out by Gravish et al. [25], the strength of a bundle of staples is also a function of the volume fraction of staples in the bundle. Following this model, we write the average entanglement density in the bundle, as the average number of entangled staples in a $l \times l \times l$ volume (i.e., the unit volume is expressed in unit of $l$):

$$\phi_t = \phi p_t \qquad (9)$$

Figure 5d shows $\phi_t$ as a function of the crown leg angle $\theta$. The packing factor $\phi$ is greater for staples with smaller $\theta$, so the effect of applying equation (9) is a slight shift of the peak entanglement from $\theta = 50°$ for $p_t$ to an optimum at $\theta = 35\text{-}55°$ for $\phi_t$. In this case, the profiles for $p_t$ and $\phi_t$ are very similar (which will not be necessarily the case when other geometrical parameters of the staples will be varied in the upcoming sections of this report). The model captures a competition of effects resulting in an optimum geometry for entanglement: (i) The density of staples is the highest for staples with low $\theta$ (Fig. 3c); (ii) Rod like staples cannot engage, and the engagement probability is the highest near 70° (Fig. 5a); (iii) if they engage, the probability of "trapping" or "entangling" other staples is higher in staples with low $\theta$. We finally note that the maximum entanglement density near $\theta$ = 35-55° (Fig. 5d) and no entanglement near $\theta$ = 0° and $\theta$ = 180° is consistent with the experimental results from the pick-up tests (Fig. 2). This result suggests that this two-staple throw-bounce-tangle model can be used to predict entanglement density in a bundle of staples, and that the entanglement density, in turn, governs the strength of the bundle. Next, we test the throw-bounce-tangle model against experiments on other types of staples.

## 4 Effects of leg length

The relative leg length $w/l$ is another critical geometrical parameter in individual staples [25]. To explore this parameter experimentally, we decreased the leg lengths of standard staples by milling the legs of sticks of staples from $w$ = 6.35 mm for standard staples down to 5.1 mm, 4.1 mm, 2.7 mm, or 1.5 mm. We also acquired other staples with longer legs ($w$ = 9.5 mm and $w$ = 12.7 mm, $d$ = 0.74 × 0.59 mm$^2$, Swingline, IL). Three pick-up tests were then performed on bundles of 1000 staples for each of these geometries. Figure 6b shows the pick-up fraction as a function of normalized leg length $w/l$. We observed a clear peak at $w/l$ = 0.4, and a sharp decrease for shorter or longer legs. This peak ratio $w/l$ is identical to the optimum found with column-collapse experiments in Gravish et al. [25].

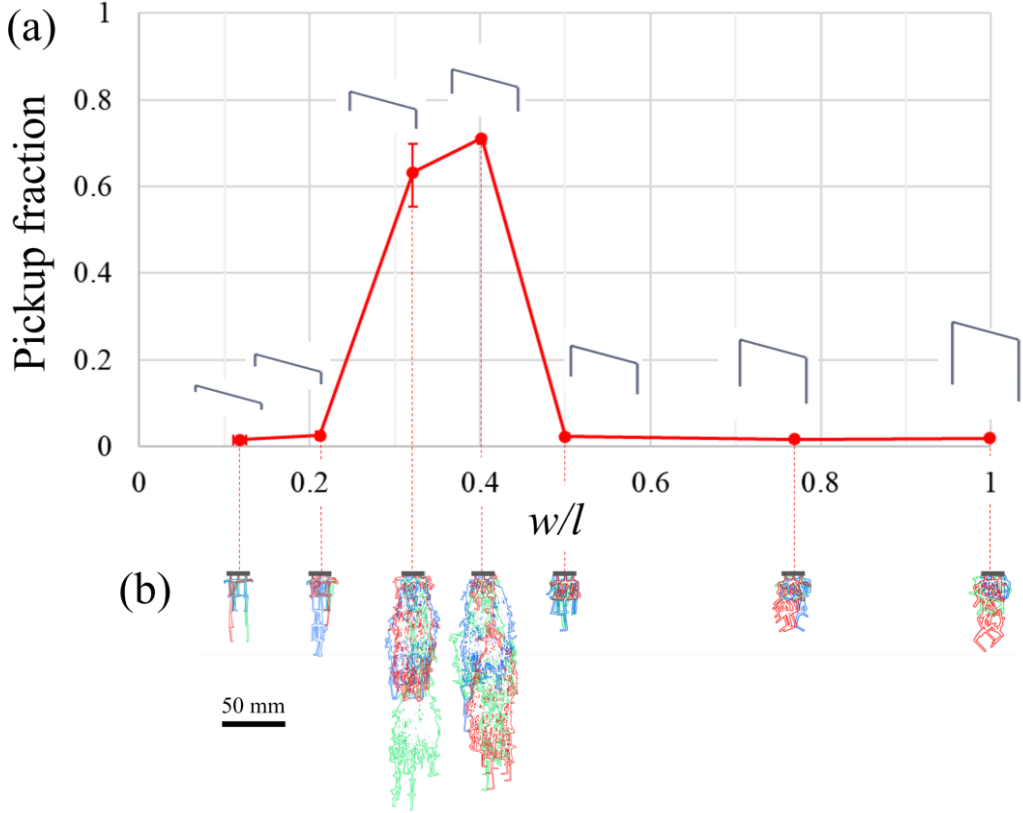

**Fig. 6** Effect of leg length: **a** Experimental pick-up fraction as a function of normalized leg length *w/l*; **b** Composite images of the picked-up bundles.

Figure 7a shows the volumetric density $\phi$ of staples measured experimentally and fitted with the model of equation (3). As expected, longer legs for the individual staples result to a rapid decrease in $\phi$. Figure 7b-f show the results of the throw-bounce-tangle model, which recovers the competing effects discussed in Gravish et al. [25]: As leg length is increased higher visibility and higher chances of engagement (Fig. 7b, c) result in a rapid increase of probability for entanglement (Fig. 7e). However longer legs also lead to a rapid decrease in $\phi$ (Fig. 7a), so that these two competing effects give rise to an optimum *w/l* value for entanglement at about *w/l* = 0.4, which is in good agreement with the optimum from the experiments and with previous results [14].

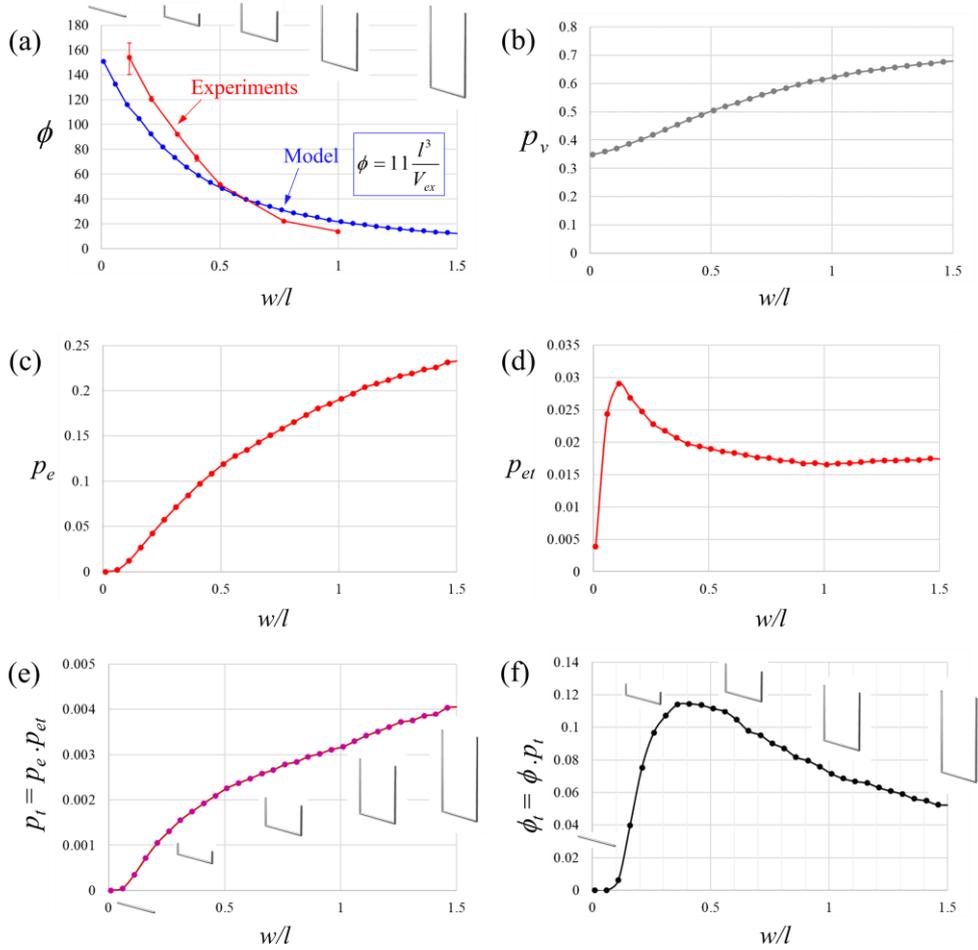

**Fig. 7** Results from the throw-bounce-tangle model for a pair of staples as a function of leg length $w/l$: **a** Experimental and modeled staple volumetric density; **b** "Visibility" $p_v$ of individual staples; **c** Probability of engagement; **d** Transition probability from engaged to entangled; **e** Overall probability of entanglement and **f** Volumetric density of entangled staples as a function of leg length $w/l$.

## 5 Effect of backbone thickness

Intuitively, we expected the thickness of the backbone of the staples to have minimal effects on entanglement as long as the backbone of the staples is "sufficiently thin" compared to its other dimensions. 3D printing of particles, which has been used in previous studies [16, 20], has relied on this assumption: Because of the limitations in printer resolution and material choice (polymers), 3D printed particles for entangles have backbones with a relative cross-section which is greater than in steel staples. However, in preliminary experiments, we noticed that 3D printed polymeric particles did not generally entangle well compared to thin steel staples. In this section, we explore the effects of backbone thickness. As a reference, we used the $\theta = 60°$ steel staple design described above, which has a backbone cross section $d^2 = 0.41 \times 0.48$ mm². We fabricated two other designs, with a backbone contour identical to the $\theta = 60°$ steel staple but with thicker backbone cross sections: $d^2 = 0.85 \times 0.85$ mm² and $d^2 = 1.50 \times 1.50$ mm². These two thicker types of staples were cut from 0.85 mm and 1.5 mm thick acrylic sheets using a precision laser cutter (Nova 35, Thunder Laser, TX). The pick-up tests revealed a sharp decrease in entanglement for thicker staples (Fig. 8): Doubling the thickness of the backbone from $d/l = 0.035$ to $d/l = 0.065$ resulted in a 75% decrease in pick-up ratio, even though the backbones in these cases may still be considered "thin" ($d/l \ll 1$).

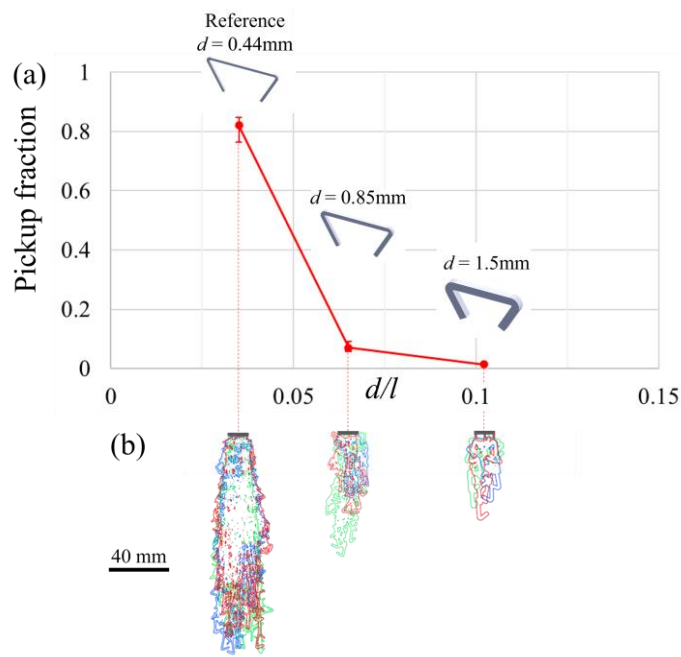

**Fig. 8** Effect of backbone thickness: **a** Experimental pick-up fraction as a function of normalized backbone thickness $d/l$; **b** Composite images of the picked-up bundles.

Figure 9 shows the results of the throw-bounce-tangle simulations. Except for the probabilities of entanglement from engagement $p_{et}$, all characteristics decrease with $d/l$. This includes the volumetric density, as well as the visibility $p_v$, which one could think would increase for thicker staples, but which decreases because thicker staples are further apart in a bundle. As a result, the entanglement density $\phi_t$ decreases very rapidly when the backbone thickness $d/l$ is increased, in accordance with the experimental results. The model therefore reveals the main two contributors of the poor entanglement of thickness staples: Decreased staple density $\phi$ and decreased probabilities of engagement $p_e$.

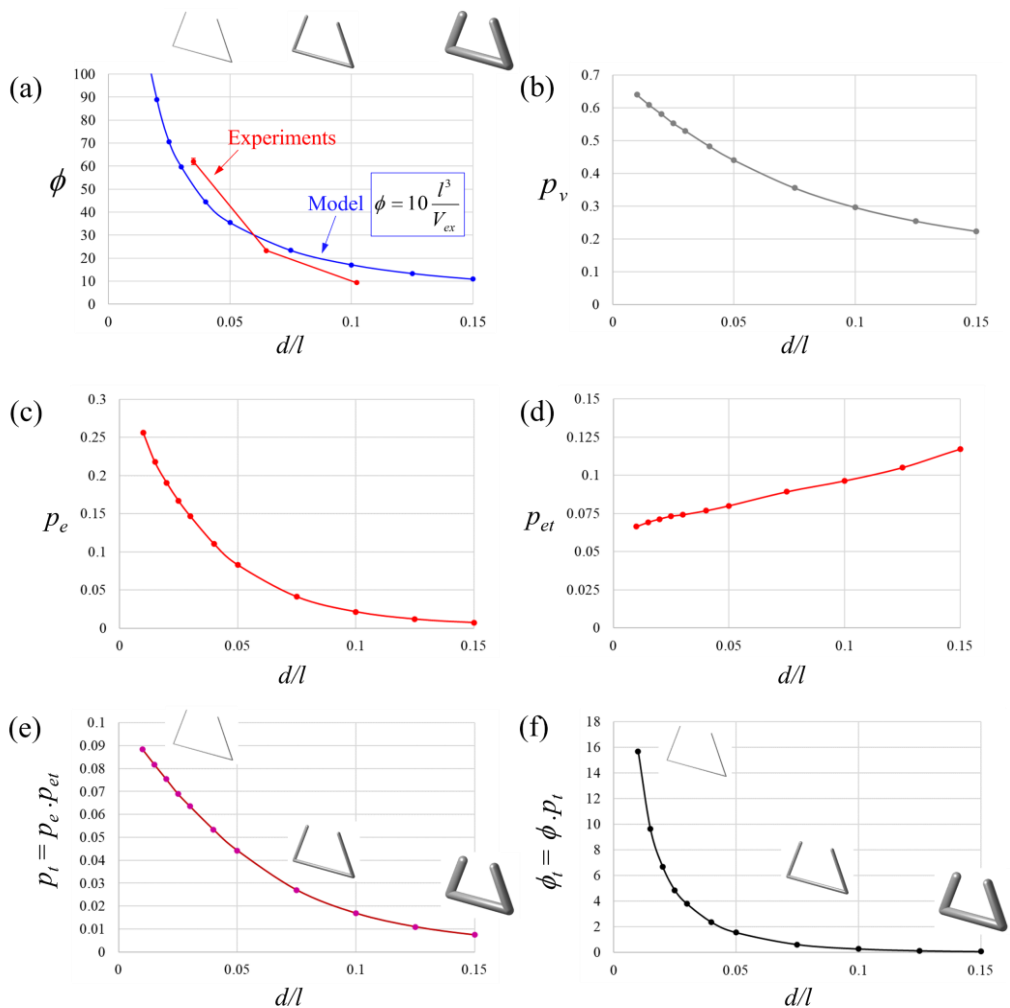

**Fig. 9** Results from the throw-bounce-tangle model for a pair of staples as a function of backbone thickness $d/l$: **a** Experimental and modeled staple volumetric density; **b** "Visibility" $p_v$ of individual staples; **c** Probability of engagement; **d** Transition probability from engaged to entangled; **e** Overall probability of entanglement and **f** Volumetric density of entangled staples as a function of backbone thickness $d/l$.

# 6 Effect of twisting

Experiments on 3D-printed staple-like particles have shown that entanglement and strength can also be manipulated by twisting the staples about the axis of the crown [20]. In this final section, we examine the effect of twisting the legs of individual staples. As a reference, we used the $\theta = 90°$ and $60°$ steel staples described above. For each of these angles, we explored two twisted designs: A set of staples twisted by an angle $\beta = 90°$ about the axis of the crown, and another set of staples twisted by an angle $\beta = 180°$ about the axis of the crown (Fig. 10a, b). Figure 10a shows the results for $\theta = 90°$ staples: Twisting angle had almost no effects on the pick-up fraction. However, any possible effect might be obscured by the already very low pick-up fraction for $\theta = 90°$ staples (0.02). The results for the twisted, $\theta = 60°$ staples (Fig. 10b) show a slightly more pronounced effect of twisting: Compared to the reference staple ($\beta = 0°$), entanglement increases by about 3.9% in staple twisted to $\beta = 90°$, but that entanglement degrades to below reference values for $\beta = 180°$. In addition, we found that the picked bundle for the $\beta = 90°$ case was more "compact" and less elongated than for $\beta = 0°$ and $\beta = 180°$, a possible indication of better entanglement stability.

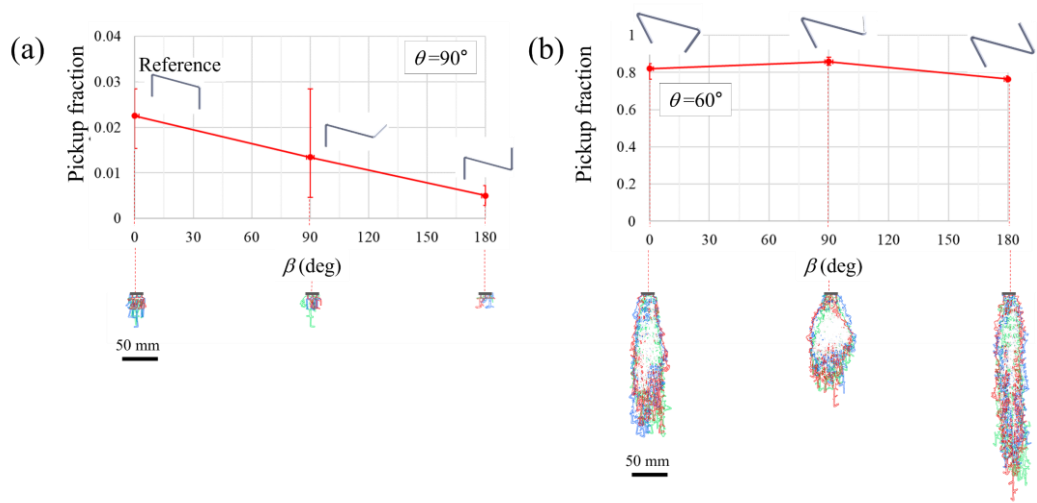

**Fig. 10** Effect of twisting staples: Experimental pick-up fraction as a function of twist angle $\beta$ for staples with **a** $\theta = 90°$ and **b** $\theta = 60°$.

Figure 11 shows the results from the throw-bounce-tangle model, which predicts a very modest effect of twist angle $\beta$ on every measure of density and probabilities. Murphy et al. [20], however, explained the entanglement of twisted staples in the context of the "pancaking" effect: Under the combined effects of vibration and gravity, the crown and the legs of the staples tend to align on horizontal planes perpendicular to gravity (Fig. 12a). In twisted staples, the legs cannot both lie on that plane, which statistically increases the probability of entanglement between staples from different horizontal planes.

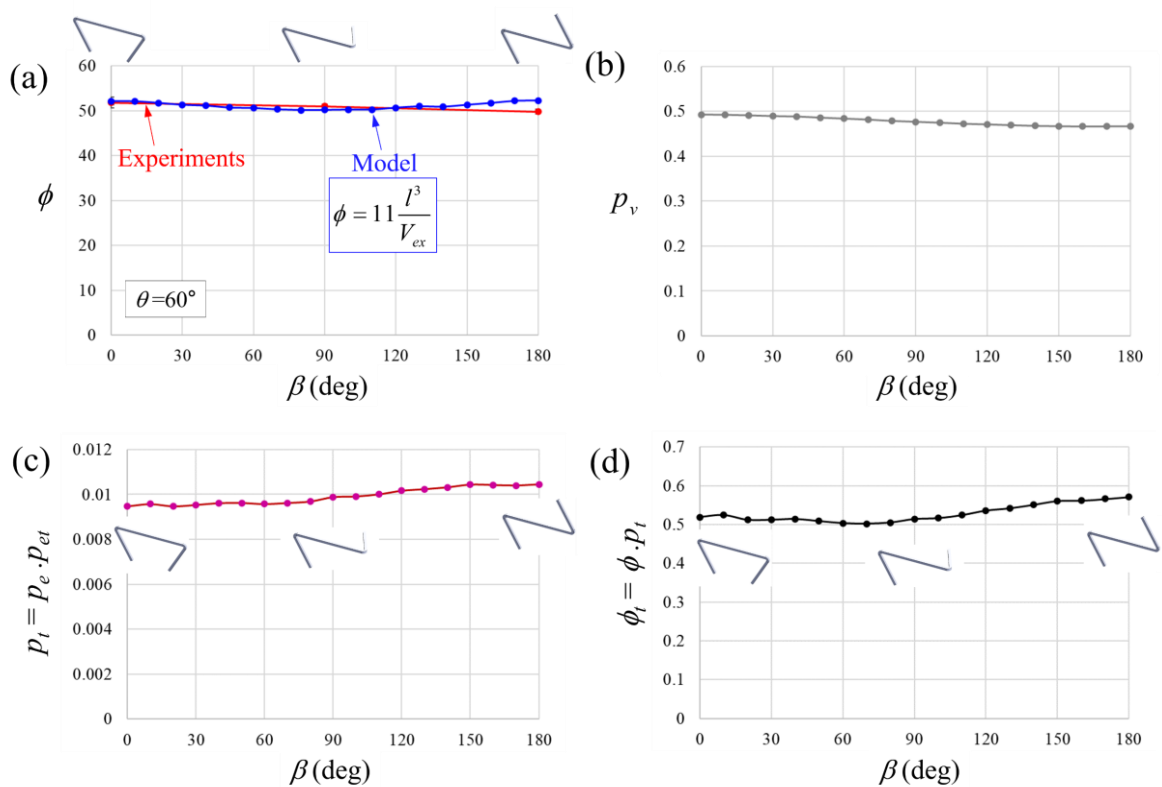

**Fig. 11** Results from the throw-bounce-tangle model for a pair of staples as a function of angle of twist $\beta$: **a** Experimental and modeled staple volumetric density; **b** "Visibility" $p_v$ of individual staples; **c** Probability of entanglement and **d** Volumetric density of entangled staples as function of angle of twist $\beta$.

We captured the "pancaking effect" effect in our model by introducing a bias on the orientation of staples 1 and 2: When we numerically created staple 1 and 2, we required that (i) the crowns of both staples 1 and 2 lie in a horizontal plane *xy* and (ii) the sum of the angles of the legs to that plane is minimized. Figure 12b shows examples of staple alignment for staples with different twist angles $\beta$. As the twist angle is increased from $\beta = 0°$ to $\beta = 90°$, the legs protrude more and more from either side of the horizontal plane. At $\beta > 90°$ there is a sharp transition, with the legs protruding from only one side of the plane (to satisfy condition (ii) above), and the out-of-plane protrusions decreasing progressively up to $\beta = 180°$. Figure 12c, d show the predictions of the throw-bounce-tangle model: Compared to the isotropic case, pancaking generally decreases the density of entanglement. However, in these pancaked staples, the density of entanglement becomes a strong function of the twist angle $\beta$. In $\beta = 90°$ and $\beta = 180°$ cases, the legs are in the *xy* plane and no engagement and entanglement are possible between staple 1 and 2. In other twisted geometries, the legs of staple 1 and 2 are out-of-plane, which greatly increased the likelihood of entanglement, with a clear maximum at $\beta = 90°$, which is consistent with our experiments and the experiments in [20]. Finally, we note that the sharp drop in entanglement predicted for $\beta > 90°$ is explained by both of the legs being on the same side of the horizontal plane, an outcome of enforcing condition (ii) above.

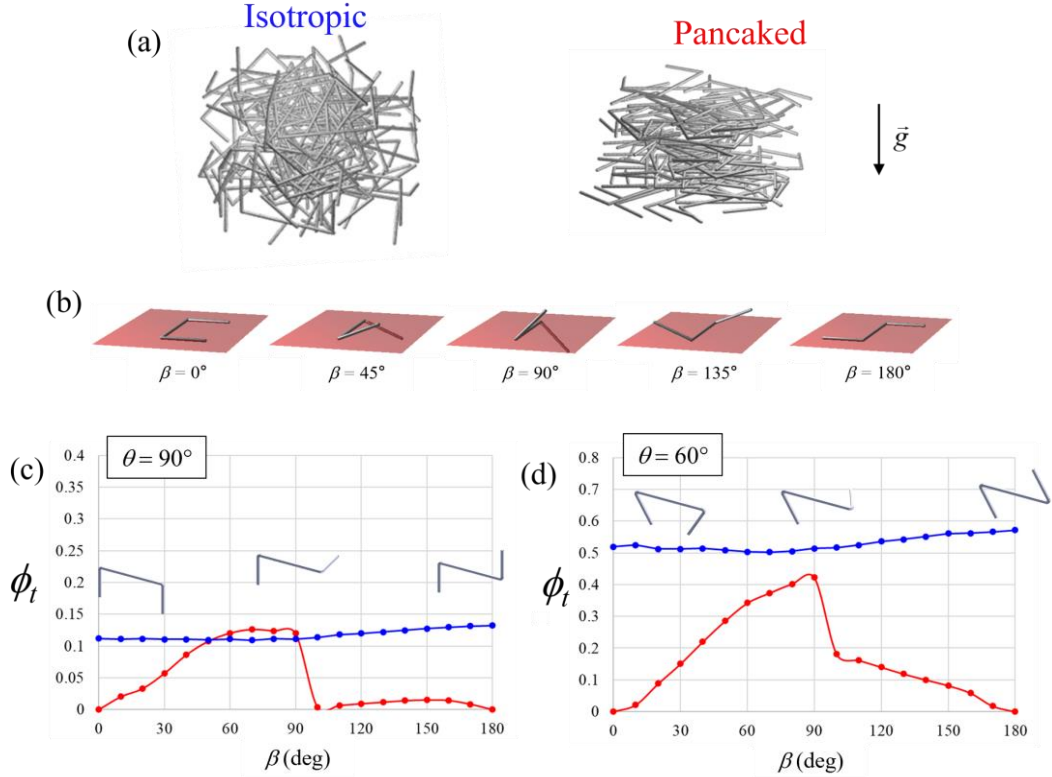

**Fig. 12** Effect of pancaking: **a** Isotropic case and pancakes cases; **b** "Pancaked" staples with various angles of twist $\beta$; Density of entanglement as a function of twist angle $\beta$ for **c** staples with $\theta = 90°$ and **d** staples with $\theta = 60°$.

## 7 An optimum staple geometry

The sections above have shown that the throw-bounce-tangle model can capture variations of entanglement for different staple geometries, in ways which are consistent with experiments and with previous studies. Because this model is relatively simple and computationally efficient, it is amenable to the exploration of large numbers of possible staple designs. In this section, we give an example where we explore the crown leg angle–leg length ($\theta$ - $w/l$) design space. We considered entanglement density over a range of $20° < \theta < 120°$ and $0 < w/l < 1.5$ for a "brute force" exploration of the design space. The combinations of these parameters that would lead to the legs of the staples crisscrossing (for which $2(w/l)\cos\theta > 1$) were excluded from this exploration.

Figure 13 shows a map of entanglement density as a function of $\theta$ and $w/l$, based on ~5000 combinations of these parameters. The map reveals a single maximum entanglement density for $\theta = 43°$ and $w/l = 0.43$, but the landscape in the design space is quite smooth, and that optimum peak is not particularly sharp. This optimum geometry is very close to the staples that we have presented earlier in that report (Fig. 2). Indeed, the experimental pick-up ratio for that design ($\theta = 45°$ and $w/l = 0.5$) was 0.8, which is the highest we measured in this study.

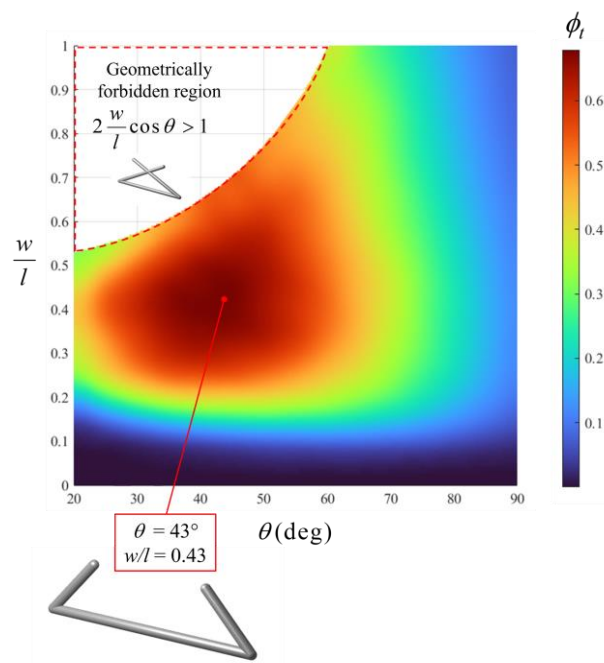

**Fig. 13** Map of entanglement density as a function of crown leg angle and leg length. The map reveals an optimum design at $\theta = 43°$ and $w/l = 0.43$.

## 8 Conclusions

Entangled matter provides intriguing perspectives in terms of deformation mechanisms, mechanical properties, assembly, and disassembly. Collective entanglement mechanisms are complex, occur over multiple length scales, and they are not fully understood to this day. In this report, we propose a simple pick-up test which can be used as a metric for entanglement. This method produces results which are consistent with other, more complicated testing methods for entanglement [15, 20, 25]. We also presented a new "throw-bounce-tangle" model, which is based on a 3D geometrical entanglement criterion between two staples, and a Monte Carlo numerical approach to extract probabilities of entanglement. This relatively simple model predicts an average density of entanglement, and it recovers the trends and optimum staple geometries identified experimentally: Entanglement can be manipulated by tuning the leg-to-crown ratio, the crown-lag angle, and the twist angle. We also show, using experiments and our model, that entanglement is very sensitive to the thickness of the backbone of the staples, even in regimes where that thickness is a small fraction (<0.04) of the other dimensions. An interesting use for this model is the design of staple-like particles optimized for entanglement. We demonstrate that the crown leg angle – leg length design space can be explored exhaustively using brute force to identify optimum particle designs. Because the throw-bounce-tangle is computationally highly efficient, it can be used in the future to explore more complex three-dimensional designs (for example, branched particles with barbs). Geometric entanglement can also be exploited in microscale [27] or even nanoscale particles [28]. Interestingly, the entangled matter is prominent in nature, in the form of "passive entanglement" in bird nests [29, 30], beaver dams [31, 32], seed barbs and hooks attachments [33, 34] and also "active entanglement" in ant rafts [35, 36] or worm blobs [37]. These systems could, therefore, provide a rich source of inspiration for new entangled materials. New optimized designs for individual particles may produce entangled bundles with attractive combinations of strength, extensibility, and toughness that may soon outperform lightweight engineering materials such as solid foams and lattices.

**Acknowledgments** This work was supported by the US National Science Foundation (Mechanics and Materials and Structures, Award No. 2033991). YS was also partially supported by the Department of Mechanical Engineering at the University of Colorado.

**Author contributions** Conceptualization: YS, SP, FB; Investigation: YS; Methodology: YS, FB; Writing-original draft: YS; Writing-review and editing: YS, SP, FB; Formal Analysis: YS, FB; Supervision: FB.

# References

1. Behringer, R.P., Chakraborty, B.: The physics of jamming for granular materials: A review. Rep. Prog. Phys. (2018). https://doi.org/10.1088/1361-6633/aadc3c
2. Jaeger, H.M.: Celebrating soft matter's 10th anniversary: Toward jamming by design. Soft Matter (2015). https://doi.org/10.1039/C4SM01923G
3. Murphy, K.A., MacKeith, A.K., Roth, L.K., Jaeger, H.M.: The intertwined roles of particle shape and surface roughness in controlling the shear strength of a granular material. Granular Matter (2019). https://doi.org/10.1007/s10035-019-0913-7
4. Gans, A., Pouliquen, O., Nicolas, M.: Cohesion-controlled granular material. Phys. Rev. E (2020). https://doi.org/10.1103/PhysRevE.101.032904
5. Karuriya, A.N., Simoes, J., Barthelat, F.: Fully dense and cohesive fcc granular crystals. Extreme Mechanics Letters (2024). https://doi.org/10.1016/j.eml.2024.102208
6. Philipse, A.P.: The random contact equation and its implications for (colloidal) rods in packings, suspensions, and anisotropic powders. Langmuir (1996). https://doi.org/10.1021/la950671o
7. Wouterse, A., Luding, S., Philipse, A.P.: On contact numbers in random rod packings. Granular Matter (2009). https://doi.org/10.1007/s10035-009-0126-6
8. Blouwolff, J., Fraden, S.: The coordination number of granular cylinders. Europhys. Lett. (2006). https://doi.org/10.1209/epl/i2006-10376-1
9. Jung, Y., Plumb-Reyes, T., Lin, H.-Y.G., Mahadevan, L.: Entanglement transition in random rod packings. arXiv (2023). https://doi.org/10.48550/arXiv.2310.04903
10. Barés, J., Zhao, Y., Renouf, M., Dierichs, K., Behringer, R.: Structure of hexapod 3d packings: Understanding the global stability from the local organization. EPJ Web Conf. (2017). https://doi.org/10.1051/epjconf/201714006021
11. Aponte, D., Estrada, N., Barés, J., Renouf, M., Azéma, E.: Geometric cohesion in two-dimensional systems composed of star-shaped particles. Phys. Rev. E (2024). https://doi.org/10.1103/PhysRevE.109.044908
12. Cantor, D., Cárdenas-Barrantes, M., Orozco, L.F.: Bespoke particle shapes in granular matter. Pap. Phys. (2022). https://doi.org/10.4279/pip.140007
13. Dierichs, K., Menges, A.: Towards an aggregate architecture: Designed granular systems as programmable matter in architecture. Granular Matter (2016). https://doi.org/10.1007/s10035-016-0631-3
14. Gravish, N., Franklin, S.V., Hu, D.L., Goldman, D.I.: Entangled granular media. Phys. Rev. Lett. (2012). https://doi.org/10.1103/PhysRevLett.108.208001
15. Franklin, S.V.: Extensional rheology of entangled granular materials. Europhys. Lett. (2014).

https://doi.org/10.1209/0295-5075/106/58004
16. Karapiperis, K., Monfared, S., Macedo, R.B.d., Richardson, S., Andrade, J.E.: Stress transmission in entangled granular structures. Granular Matter (2022). https://doi.org/10.1007/s10035-022-01252-4
17. Marschall, T.A., Franklin, S.V., Teitel, S.: Compression-and shear-driven jamming of u-shaped particles in two dimensions. Granular Matter (2015). https://doi.org/10.1007/s10035-014-0540-2
18. Zhao, Y., Barés, J., Socolar, J.E.: Yielding, rigidity, and tensile stress in sheared columns of hexapod granules. Phys. Rev. E (2020). https://doi.org/10.1103/PhysRevE.101.062903
19. Pezeshki, S., Sohn, Y., Fouquet, V., Barthelat, F.: Tunable entanglement and strength in "granular metamaterials" based on staple-like particles: Experiments and discrete element models. In review (2024)
20. Murphy, K.A., Reiser, N., Choksy, D., Singer, C.E., Jaeger, H.M.: Freestanding loadbearing structures with z-shaped particles. Granular Matter (2016). https://doi.org/10.1007/s10035-015-0600-2
21. Murphy, K., Roth, L., Peterman, D., Jaeger, H.: Aleatory construction based on jamming: Stability through self-confinement. Archit. Des. (2017). https://doi.org/10.1002/ad.2198
22. Trepanier, M., Franklin, S.V.: Column collapse of granular rods. Physical Review E—Statistical, Nonlinear, and Soft Matter Physics (2010). https://doi.org/10.1103/PhysRevE.82.011308
23. Govender, N., Wilke, D.N., Wu, C.-Y., Khinast, J., Pizette, P., Xu, W.: Hopper flow of irregularly shaped particles (non-convex polyhedra): Gpu-based dem simulation and experimental validation. Chem. Eng. Sci. (2018). https://doi.org/10.1016/j.ces.2018.05.011
24. Li, L., Marteau, E., Andrade, J.E.: Capturing the inter-particle force distribution in granular material using ls-dem. Granular Matter (2019). https://doi.org/10.1007/s10035-019-0893-7
25. Gravish, N., I. Goldman, D.: Entangled granular media. In: Fernandez-Nieves, A., Puertas, A.M. (eds.) Fluids, colloids and soft materials - an introduction to soft matter physics, pp. 341-354. John Wiley & Sons, Inc, (2016).
26. TheMathWorksInc: Matlab version: 9.13.0 (r2022b). (2022). https://www.mathworks.com
27. Paulsen, K.S., Deng, Y., Chung, A.J.: Diy 3d microparticle generation from next generation optofluidic fabrication. Adv. Sci. (2018). https://doi.org/10.1002/advs.201800252
28. Miszta, K., De Graaf, J., Bertoni, G., Dorfs, D., Brescia, R., Marras, S., Ceseracciu, L., Cingolani, R., Van Roij, R., Dijkstra, M.: Hierarchical self-assembly of suspended branched colloidal nanocrystals into superlattice structures. Nat. Mater. (2011). https://doi.org/10.1038/nmat3121
29. Weiner, N., Bhosale, Y., Gazzola, M., King, H.: Mechanics of randomly packed filaments—the "bird nest" as meta-material. J. Appl. Phys. (2020). https://doi.org/10.1063/1.5132809
30. Bhosale, Y., Weiner, N., Butler, A., Kim, S.H., Gazzola, M., King, H.: Micromechanical origin of plasticity and hysteresis in nestlike packings. Phys. Rev. Lett. (2022). https://doi.org/10.1103/PhysRevLett.128.198003
31. Żurowski, W.: Building activity of beavers. Acta Theriol. (1992). https://doi.org/10.4098/AT.ARCH.92-41
32. Larsen, A., Larsen, J.R., Lane, S.N.: Dam builders and their works: Beaver influences on the structure and function of river corridor hydrology, geomorphology, biogeochemistry and ecosystems. Earth-Sci. Rev. (2021). https://doi.org/10.1016/j.earscirev.2021.103623
33. Claus, C.G., Gorb, E.V., Gorb, S.N., Li, C.: Comparative study on mechanical properties and biomineralization of hooks in the diaspores of three epizoochorous plant species. Acta Biomater. (2024). https://doi.org/10.1016/j.actbio.2024.07.041
34. Sorensen, A.E.: Seed dispersal by adhesion. Annu. Rev. Ecol. Syst. (1986). https://www.jstor.org/stable/2097004
35. Hu, D., Phonekeo, S., Altshuler, E., Brochard-Wyart, F.: Entangled active matter: From cells to ants. Eur. Phys. J.: Spec. Top. (2016). https://doi.org/10.1140/epjst/e2015-50264-4
36. Phonekeo, S., Dave, T., Kern, M., Franklin, S.V., Hu, D.L.: Ant aggregations self-heal to

compensate for the ringelmann effect. Soft Matter (2016). https://doi.org/10.1039/C6SM00063K
37. Worley, A., Sendova-Franks, A.B., Franks, N.R.: Social flocculation in plant–animal worms. R. Soc. Open Sci. (2019). https://doi.org/10.1098/rsos.181626